\documentclass[fleqn,10pt]{SelfArx}
\usepackage{lipsum}

\definecolor{color1}{RGB}{0,0,90} % Color of the article title and sections
\definecolor{color2}{RGB}{0,20,20} % Color of the boxes behind the abstract an

\usepackage{hyperref} % Required for hyperlinks
\hypersetup{hidelinks,colorlinks,breaklinks=true,urlcolor=color2,citecolor=color1,linkcolor=color1,bookmarksopen=false,pdftitle={Title},pdfauthor={Author}}

 \renewcommand{\vec}[1]{\mbox{\boldmath $#1$}}

\JournalInfo{Geomagnetism \& Aeronomy, 2016, Vol.\,56, No.\,8} % Journal information
\Archive{}
\PaperTitle{
Meridional circulation in the Sun and stars
}
\Authors{L.~L.~Kitchatinov\textsuperscript{1,2}*}
\affiliation{\textsuperscript{1}\textit{Institute for Solar-Terrestrial Physics, Lermontov Str. 126A, Irkutsk, 664033, Russia}}
\affiliation{\textsuperscript{2}\textit{
Pulkovo Astronomical Observatory, Pulkovskoe Sh. 65, St. Petersburg, 196140, Russia
}}
\affiliation{*\textbf{E-mail}: kit@iszf.irk.ru}

\Keywords{}
\newcommand{\keywordname}{Keywords}

\Abstract{
Mean-field hydrodynamics advanced to clear explanations for the origin and properties of the global meridional flow in stellar convection zones. Qualitative arguments and analysis of basic equations both show that the meridional circulation is driven by non-conservative centrifugal and buoyancy forces and results from a slight disbalance between these two drivers. The deviations from the thermal wind balance are relatively large near the boundaries of convection zones. Accordingly, the meridional flow attains its largest velocities in the boundary layers and decreases inside the convection zone. This picture, however, is neither supported nor dismissed by the conflicting results of recent helioseismic soundings or 3D numerical experiments. The relevant physics of the differential temperature and its possible relation to the solar oblateness are briefly discussed.
}

\begin{document}

\flushbottom % Makes all text pages the same height

\maketitle % Print the title and abstract box

\tableofcontents % Print the contents section

\thispagestyle{empty} % Removes page numbering from the first page

%----------------------------------------------------------------------------------------
%	ARTICLE CONTENTS
%----------------------------------------------------------------------------------------
%%%%%%%%%%%%%%%%%%%%%%%%%%%%%%%%%%%%%%%%%%%%%%%%%%%%%%%%%%%%%%%%%%%%%%%%%%%%%
\section{Introduction} % The \section*{} command stops section numbering
%%%%%%%%%%%%%%%%%%%%%%%%%%%%%%%%%%%%%%%%%%%%%%%%%%%%%%%%%%%%%%%%%%%%%%%%%%%%%
Meridional flow is important for solar magnetic activity. Dynamo theory explains the observed equatorial migration of sunspot activity by the advection of toroidal magnetic fields by the deep meridional flow (Wang et al., 1991; Choudhuri et al., 1995; Durney, 1995). The near-surface poleward flow takes part in latitudinal migration and polar reversals of the poloidal field (Sheeley, 2005).  It has been even conjectured that grand minima in solar activity, similar to the famous Maunder minimum, are caused by fluctuations in the meridional flow (Karak and Choudhuri, 2009).

This paper discusses the current theoretical understanding of the meridional flow. The flow in the Sun and solar-type stars is restricted to their convective envelopes. Its penetration into the underlying radiation zones is shallow (Gilman and Miesch, 2004; Kitchatinov and Rüdiger, 2005). More specifically, radiation cores have their own very slow rotationally induced circulation (Tassoul, 2000). The meridional flow can be defined as a poloidal part of the global axisymmetric motion resulting from an averaging - over time or longitude or ensemble of convective motions - of the velocity field. The meridional flow is therefore a natural subject for mean-field hydrodynamics. Studies of the flow have a long history. Its current understanding has advanced to the stage where clear qualitative explanations can be given for its origin and properties. The discussion to follow will, nevertheless, involve the basic equations and the results of numerical mean-field models to support the pictorial argumentation. Fragmental discussions of 3D numerical experiments are, however, far from complete. The meridional flow theory will be confronted with recent helioseismic inversions.
%%%%%%%%%%%%%%%%%%%%%%%%%%%%%%%%%%%%%%%%%%%%%%%%%%%%%%%%%%%%%%%%%%%%%%%%%%%%%
\section{Two main sources of the meridional flow}
%%%%%%%%%%%%%%%%%%%%%%%%%%%%%%%%%%%%%%%%%%%%%%%%%%%%%%%%%%%%%%%%%%%%%%%%%%%%%
The mean meridional flow in a stellar convection zone is subject to a drag by turbulent viscosity. Some drivers supplying energy to the meridional flow are, therefore, necessary to support the flow against viscous decay.

An important conclusion about meridional flow driving follows from the seemingly trivial fact that the fluid particles in a (steady) flow circulate over closed trajectories. Strictly speaking, the trajectories are not closed because of the non-uniform rotation, but azimuthal displacements play no role for statistically axisymmetric background state. It is known from classical mechanics that only non-conservative forces can transmit energy to particles circulating on closed trajectories. The basic motion equation is, therefore, not convenient for describing meridional circulation. The equation contains ‘well-disguised zeros’ – the potential forces. These conservative forces enter the equation with large coefficients, which makes the equation difficult to solve, especially when it is solved numerically. The standard recipe for avoiding this difficulty is to curl the equation to filter-out the potential forces. The resulting vorticity equation,
\begin{eqnarray}
    \frac{\partial\omega}{\partial t}
    &+& r\sin\theta{\vec\nabla}\cdot\left({\vec V}^\mathrm{m}\frac{\omega}{r\sin\theta}\right)\
    +\ {\cal D}\left(\vec{V}^\mathrm{m}\right)\ =
    \nonumber \\
    &=&\sin\theta\ r\frac{\partial\Omega^2}{\partial z}
    - \frac{g}{c_\mathrm{p} r}\frac{\partial S}{\partial\theta} ,
    \label{1}
\end{eqnarray}
contains two principal drivers of the meridional flow on its right-hand side. In this equation, the standard spherical coordinates ($r,\theta,\phi$) are used, $\vec{V}^\mathrm{m}$ is the meridional flow velocity, $\omega=(\vec{\nabla}\times\vec{V}^\mathrm{m})_\phi$  is the azimuthal vorticity, $g$ is gravity, $S$ is the specific entropy, and $\partial/\partial z = \cos\theta\partial/\partial r - \sin\theta r^{-1}\partial/\partial\theta$ is the gradient along the rotation axis. The term ${\cal D}(\vec{V}^\mathrm{m})$ in Eq. (\ref{1}) stands for the meridional flow damping by the eddy viscosity. The explicit expression for this term is rather bulky (its relation to the eddy viscosity tensor can be found in Kitchatinov and Olemskoy,  2011).

It may be noted that equation (\ref{1}) results from curling the motion equation for unit mass, $\partial{\vec V}/\partial t + ({\vec V}\cdot{\vec\nabla})V + ...$. The equally valid equation for unit volume, $\partial (\rho V_i )/\partial t + \nabla_j (\rho V_i V_j ) + ...$, is not appropriate because the fluid particles circulating on closed trajectories with the meridional flow conserve mass, not volume.

The first term on the right-hand side of (\ref{1}) stands for the driving of the meridional flow by centrifugal force. This force is known to be conservative if the angular velocity does not depend on the cylindrical coordinate $z = r \cos\theta$, i.e., if $\Omega$ is constant on the cylinders. Accordingly, the centrifugal driving of the meridional flow is proportional to the $z$-gradient of $\Omega$. Centrifugal excitation of the meridional flow is long known. Einstein (1926) explained the shape of watercourse cross-sections at river turns by this effect.  As the main stream at the river turn is similar to non-uniform rotation, Einstein concluded that there must be a circulation flow across the main stream with the surface flow towards the external side of the turn and the near-bottom flow towards its inner side. Kieppenhahn (1963) was probably the first to discuss centrifugal excitation of the meridional flow in stars. This effect is now named ‘gyroscopic pumping’ (Garaud and Bodenheimer, 2010).

\begin{figure}[thb]
\includegraphics[width= \linewidth]{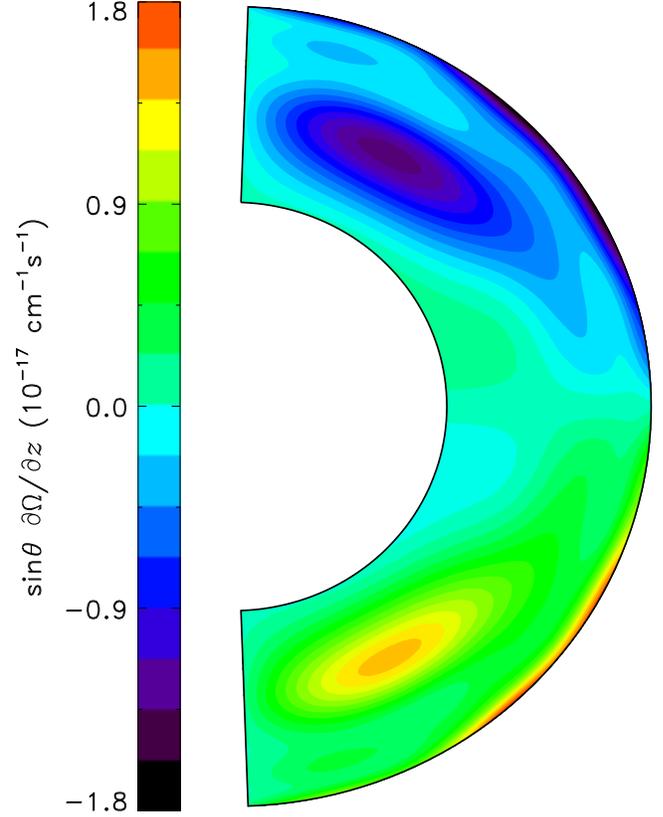}
\\
\caption{Distribution of the quantity $\sin\theta\partial\Omega/\partial z$
        of the gyroscopic pumping term of Eq.\,(\ref{1}) inside the Sun
        based on the helioseismic data of Antia et al. (2008). The generated
        vorticity is negative in the northern hemisphere (anti-clockwise
        circulation) and positive in the southern hemisphere.}
    \label{f1}
\end{figure}

The angular velocity gradients in the solar interior were detected seismologically by Antia et al. (2008). Figure 1 shows the distribution of the gyroscopic pumping term of Eq.\,(1) inside the Sun constructed using their data ({\sl GONG} data averaged over the 23rd activity cycle). The gyroscopic pumping generates anti-clockwise circulation in the northern hemisphere (in the western part of the meridional cross-section of the convection zone) and clockwise circulation in the southern hemisphere.

For incompressible fluids with uniform density, gyroscopic pumping is the only principal driver of the meridional flow. The gaseous material of stellar convection zones is however compressible. The baroclinic driver due to the non-conservative part of the pressure force,
\begin{equation}
    -{\vec\nabla}\times\frac{1}{\rho}{\vec\nabla}P =
    \frac{1}{\rho^2}{\vec\nabla}\rho\times{\vec\nabla}P =
    - \frac{1}{c_\mathrm{p}\rho}{\vec\nabla}S\times{\vec\nabla}P ,
    \nonumber
\end{equation}
is equally significant. This baroclinic driver is accounted for by the second term on the right-hand side of Eq.\,(\ref{1}). It is related to the ‘differential temperature’ – the temperature difference between the equator and poles.
%%%%%%%%%%%%%%%%%%%%%%%%%%%%%%%%%%%%%%%%%%%%%%%%%%%%%%%%%%%%%%%%%%%%%%%%%%%%%
\section{Differential temperature}
%%%%%%%%%%%%%%%%%%%%%%%%%%%%%%%%%%%%%%%%%%%%%%%%%%%%%%%%%%%%%%%%%%%%%%%%%%%%%
There have been multiple attempts to observe the equator-to-pole temperature difference on the Sun motivated primarily by the differential rotation theory. Measurements by Rast et al. (2008) suggest that the poles are warmer than the equator by about 2.5 K.

The differential temperature can be understood as a consequence of the influence of rotation on thermal convection. Weiss (1965) noticed that the eddy thermal diffusivity should increase from the equator to the poles under the influence of rotation. The higher polar diffusivity results in warmer poles. The buoyancy force then tends to produce a clock-wise meridional flow (in the north-west quadrant of the meridional cross-section).

The meridional flow can in turn transport angular momentum and produce differential rotation. It is instructive to consider this type of angular momentum transport because the anti-solar rotation in recent 3D simulations (e.g., Matt et al., 2011; Guerrero et al., 2013;  Karak et al. 2015) most probably results from this effect.

\begin{figure}[thb]
\centering
\includegraphics[width= 7 truecm]{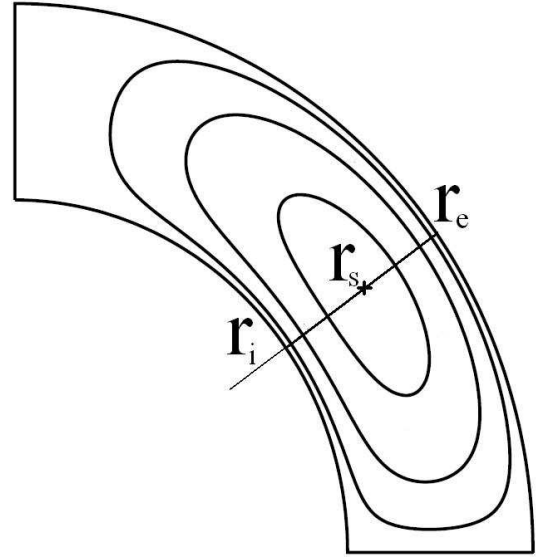}
\\[0.3 truecm]
\caption{Full lines show the meridional flow stream-lines in a convective
        envelope between the inner radius $r_\mathrm{i}$ and the external
        radius $r_\mathrm{e}$. $r_\mathrm{s}$ is the radius of the stagnation
        point where meridional velocity falls to zero. The direction of the angular momentum flux across the conical surface shown by the straight radial line coincides with the direction of the surface meridional flow.}
    \label{f2}
\end{figure}

Angular momentum flux across a conical surface of a certain $\theta$ value reads
\begin{equation}
    M = 2\pi\sin^3\theta \int\limits_{r_\mathrm{i}}^{r_\mathrm{e}}
    \rho V_\theta^\mathrm{m}\Omega r^3\mathrm{d}r ,
    \label{2}
\end{equation}
where $V_\theta^\mathrm{m}$  is the meridional flow velocity and $r_\mathrm{i}$ and $r_\mathrm{e}$ are the radii of the internal and external boundaries of the convection zone respectively (Fig.\,\ref{f2}). The mass flux across the same conical surface should be zero:
\begin{equation}
    \int\limits_{r_\mathrm{i}}^{r_\mathrm{e}}
    \rho V_\theta^\mathrm{m} r\mathrm{d}r = 0\ .
    \label{3}
\end{equation}
Differential rotation is small for the slow rotation case, for which anti-solar rotation is met in the simulations. The angular velocity in Eq.\,(\ref{2}) can be taken outside the integral sign in this case to rewrite the angular momentum flux (\ref{2}) as follows,
\begin{equation}
    M = 2\pi\sin^3\theta\ \Omega\int\limits_{r_\mathrm{i}}^{r_\mathrm{e}}
    \rho V_\theta^\mathrm{m}\Omega r (r^2 - r_\mathrm{s}^2)\mathrm{d}r ,
    \label{4}
\end{equation}
where $r_\mathrm{s}$ is the stagnation point radius (Fig.\,\ref{f2}) and the mass conservation of Eq.\,(\ref{3}) is accounted for. This expression is sign-definite. The sense of the latitudinal flux (\ref{4}) of angular momentum coincides with the sense of the surface meridional flow (Kitchatinov, 2006). The differential temperature in 3D simulations is usually suppressed by a prescribed large isotropic thermal diffusion. Then, the meridional flow is dri\-ven by gyroscopic pumping alone. Convective transport of angular momentum is downward in the case of slow rotation (K\"{a}pyl\"{a} et al., 2011; R\"{u}diger et al., 2013). The resulting increase in angular velocity with depth produces meridional circulation with poleward surface flow via the gyroscopic pumping effect. The anti-solar rotation results in this case in accord with equation (\ref{4}).

This scenario does not apply to mean-field models because the thermal diffusivity of the models includes rotationally induced anisotropy. The eddy thermal diffusion along the rotation axis is larger compared to the direction normal to this axis. The anisotropy in particular means that the eddy heat flux and the entropy gradient are not strictly aligned and a poleward meridional flux of heat is present in the rotating convection zone even if the entropy gradient is almost radial. The meridional flux is more efficient by far in producing the differential temperature than a latitude-dependent but isotropic thermal diffusion (Rüdiger et al., 2005). The differential temperature is equally important in driving the meridional flow compared with the gyroscopic pumping even in the case of slow rotation. Only if the heat transport anisotropy is artificially suppressed, can the anti-solar rotation be found (Kitchatinov and Olemskoy, 2012).

The pole-equator temperature difference computed with a mean-field model (combined computation of the differential rotation, meridional flow and heat transport) is shown in Fig.3 as the function of the radius inside the solar convection zone. The model is a slightly elaborated version of the model by Kitchatinov and Olemskoy (2011). The positive differential temperature decreases with radius and falls below 1K at the external boundary of the computation domain, which is placed shortly below the photosphere. This small surface value is probably below the resolution limit of current observations.

\begin{figure}[thb]
\includegraphics[width= \linewidth]{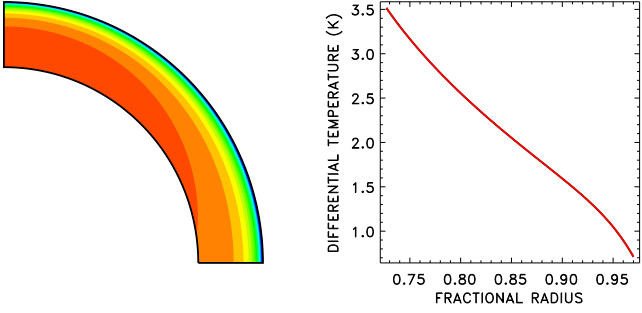}
\\
\caption{Isentropic lines in the meridional cross-section of the solar
        convection zone ({\it left panel}) and the pole-equator temperature
        difference as the function of fractional radius ({\it right panel})
        from a mean-field model computation.}
    \label{f3}
\end{figure}

The differential temperature induced by anisotropic convection can be relevant to the subtle problem of solar oblateness. The oblateness – the relative difference between the equatorial and polar solar radii – is mainly caused by solar rotation and can be estimated by the ratio of the centrifugal to gravity forces ($\sim 10^{-5}$).   This small oblateness has been formerly a topic of high interest because of its relation to the hypothetical rapid rotation in the solar core and to the theory of gravity (cf., e.g., Dicke, 1970). Interest faded after seismological detection of the almost uniform rotation of the core. However, measurements of the oblateness have reached very high precision and have provided a puzzling result: the residual oblateness after the subtraction of its centrifugal part is negative (Kuhn et al., 2012; Gough, 2015). Filtering out centrifugal oblateness leaves a prolate Sun. The origin of residual prolateness is uncertain. It could be caused by a strong ($\sim 10^8$\,Gs) internal magnetic field of the radiation core (Gough, 2015). A more routine explanation might be the effect of differential temperature.  The entropy iso-surfaces of Fig.\,\ref{f3} are prolate and the relative magnitude of the differential temperature $\delta T/T \sim 10^{-6}$ is of the same order as the residual prolateness.
%%%%%%%%%%%%%%%%%%%%%%%%%%%%%%%%%%%%%%%%%%%%%%%%%%%%%%%%%%%%%%%%%%%%%%%%%%%%%
\section{Thermal wind balance: maintenance and violation}
%%%%%%%%%%%%%%%%%%%%%%%%%%%%%%%%%%%%%%%%%%%%%%%%%%%%%%%%%%%%%%%%%%%%%%%%%%%%%
The two sources of the meridional flow on the right-hand side of Eq.\,(\ref{1}) are opposite in sign. The gyroscopic pumping produces anti-clockwise circulation (Fig.\,\ref{f1}).  The positive differential temperature of Fig.\,\ref{f3} tends to produce circulation of the opposite sense. The two sources are, therefore, competing in the solar convection zone. This is probably the case with all convective stars whose rotation is not too slow.

Eq.\,(\ref{1}) can be normalized to dimensionless units by multiplying it by $(R^2/\nu_{_\mathrm{T}})^2$; $R$ is the stellar radius and $\nu_{_\mathrm{T}}$ is the eddy viscosity. The normalization shows that the characteristic value of each of the two terms on the right-hand side of this equation is large compared to each term on the left-hand side (Kitchatinov, 2013). The two sources of the meridional flow have to almost balance each other in order to satisfy the equation. The convection zone is therefore close to the thermal wind balance:
\begin{equation}
    \sin\theta\ r\frac{\partial\Omega^2}{\partial z} =
    \frac{g}{c_\mathrm{p}r}\frac{\partial S}{\partial\theta}\ .
    \label{5}
\end{equation}
Various forms of the balance condition (\ref{5}) are widely used in astro- and geophysical fluid dynamics (Tassoul, 2000). The equation shows in particular that the solar-type rotation law with considerable deviations of isorotaion surfaces from cylinders can be reproduced only with allowance for the differential temperature (Kitchatinov and R\"{u}diger, 1995; Miesch et al., 2006; Warnecke et al., 2013).

If the balance condition (\ref{5}) is satisfied strictly, the meridional flow of Eq.\,(\ref{1}) has no sources and falls to zero. The flow, therefore, results from deviations from the thermal wind balance. Moreover, the balance is maintained by the meridional flow: deviations from the balance excite the meridional circulation, which tends to restore the balance by transporting angular momentum and heat. It is not possible, therefore, to construct a correct model or theory for the meridional flow alone. An attempt to define the meridional flow from Eq.\,(\ref{1}) alone, with its right-hand side prescribed, leads to a typical ill-posed problem. A small error in the prescribed differential rotation or temperature results in a large error in the computed circulation. Sufficient data for a reliable prescription of the internal differential temperature are lacking. The differential rotation, meridional flow and heat transport are dynamically coupled and are modelled by solving the corresponding three coupled equations together. The models are usually called \lq differential rotation models’ because non-uniform rotation is considered to be their most significant outcome.

\begin{figure}[thb]
\includegraphics[width= \linewidth]{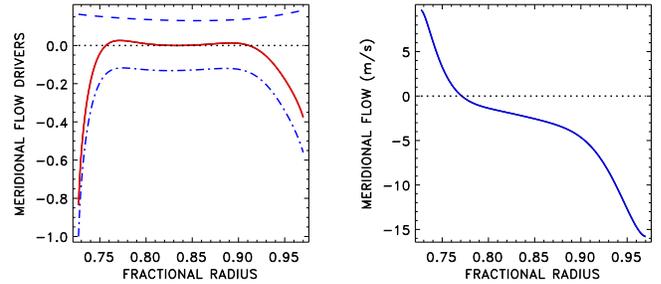}
\\
\caption{{\it Left panel}: depth profiles of the meridional flow drivers
        for the 45$^\circ$ latitude (arbitrary units). The dashed-dotted line shows the gyroscopic pumping term of Eq.\,(\ref{1}) and dashed line – the baroclinic term. The sum of the two drivers of the meridional flow is shown by the full line. The bulk of the convection zone is close to the thermal wind balance. {\it Right panel}: depth profile of the meridional velocity for the same latitude. Negative values mean poleward flow.}
    \label{f4}
\end{figure}

Figure~\ref{f4} shows the depth profiles of the meridional flow drivers and the resulting flow velocity computed with the same model as Fig.\,\ref{f3}. According to this Figure, the bulk of the convection zone is close to the thermal wind balance of Eq.\,(\ref{5}). The balance is, however, violated near the convection zone boundaries. The models naturally use a complete set of boundary conditions. An addition of an extra condition (\ref{5}) to this set would form an excessive set of conditions, which could not be satisfied all together. In other words, the thermal wind balance is not compatible with the boundary conditions. As a result, layers with thickness estimated by the Ekman depth  $D \simeq \sqrt{\nu_{_\mathrm{T}}/\Omega}$ are formed near the boundaries where balance condition (\ref{5}) is violated (Durney, 1989). The boundary layers are clearly visible in Fig.\,\ref{f4}. As the deviation from the balance gives rise to the meridional flow, the flow velocity is relatively large near the boundaries and decreases inside the convection zone.

The boundary layers of Fig.\,\ref{f4} are dominated by the negative gyroscopic pumping (Fig.\,\ref{f1}). The resulting negative azimuthal vorticity corresponds to the anti-clockwise circulation. The bottom layer is thinner because of a decrease in the eddy viscosity near the bottom. The layers thickness in the solar model $D \sim 30$\,Mm is relatively large compared to thinner layers in faster rotating stars (Kitchatinov and Olemskoy 2012).

The thermal wind balance in the bulk of the convection zone holds on average only. Fluctuations in the convective sources of the differential rotation (the $\Lambda$-effect; R\"{u}diger, 1989) produce short-term fluctuations in the angular velocity and, therefore, in the gyroscopic pumping. Rempel (2005) developed a differential rotation model with allowance for the fluctuations. The relative magnitude of the meridional flow fluctuations in his model was larger than the fluctuations in the rotation rate. Even a small fluctuating deviation from the thermal wind balance produces a considerable response in the meridional flow.
%%%%%%%%%%%%%%%%%%%%%%%%%%%%%%%%%%%%%%%%%%%%%%%%%%%%%%%%%%%%%%%%%%%%%%%%%%%%%
\section{Meridional flow structure}
%%%%%%%%%%%%%%%%%%%%%%%%%%%%%%%%%%%%%%%%%%%%%%%%%%%%%%%%%%%%%%%%%%%%%%%%%%%%%
Though the physics of the meridional flow in stellar convection zones seems to be well understood, the theoretical gro\-unds remain shaky until supported by observations. The poleward surface flow is well established by observations (Komm et al., 1993; Hathaway and Rightmire, 2010). The dependence of the flow on radius is less certain. Time-distance helioseismology (Duvall et al., 1993) remains the principal tool for measuring the dependence. Giles et al. (1997) found that the poleward flow survives up to about 28 Mm below the photosphere.

Sounding of slow meridional circulation in larger depths is a formidable task. Only recently, the first detections of the deep flow were reported and their results have not converged yet to a coherent picture. Zhao et al. (2013) found that the deep flow has a complicated structure varying on relatively small scales with radius and latitude. They detected the stagnation point, around which the flow changes direction, at a relatively small depth of about 65 Mm. The shallow reversal of the flow disagrees with the theoretical Fig. 4. Zhao et al. concluded that the flow has a double-cell structure in radius with poleward flow at the top and bottom of the convection zone and an equatorward flow in between. Shad et al. (2013) applied another method to {\sl MDI/SOHO} data for 2004-2010 and found a flow structure which differs in details from Zhao et al. (2013) but confirmed the fine flow structure with multiple cells in latitude and radius. Jackiewicz et al. (2015) analyzed two years of {\sl GONG} data and confirmed the shallow flow reversal but found a single global flow cell. They noticed a problem consisting in violation of the conservation of mass. The amount of mass carried by the detected flow from the equator to the poles is not balanced by the mass carried by the return flow.

The standard recipe of hydrodynamics for ensuring mass conservation is to specify the flow in terms of the stream function $\psi$
\begin{equation}
    {\vec V}^\mathrm{m} = \frac{1}{\rho}{\vec\nabla}\times
    \left({\vec e}_\phi\frac{\psi}{r\sin\theta}\right)\ .
    \label{6}
\end{equation}
Isolines of $\psi$ show the stream-lines of the mass-conserving meridional circulation. Equation (\ref{6}) in particular shows that the stream-function defines the complete vector of the circulation velocity so that the radial motion can be detected in line with the meridional flow. Rajaguru and Antia (2015) performed helioseismic inversions for the meridional flow in terms of the stream-function. They used four years of {\sl HMI/SDO}  data and found a single cell circulation with a deep-seated stagnation point similar to that in Fig.\,\ref{f4}. Further refinements of helioseismic detections can be anticipated.

Numerical experiments do not provide a certain conclusion either. Featherstone and Miesch (2015) found a multi-cell circulation in the solar model consistent with the helioseismic inversions of Zhao et al. (2013). Passos et al. (2015) obtained similar results but concluded that an equatorward flow should be present at the base of the convection zone. The flux-transport models for the solar dynamo are not very sensitive to the flow direction in the bulk of the convection zone but are vitally dependent on the presence of the equatorward flow at its base (Hazra et al., 2014).
%%%%%%%%%%%%%%%%%%%%%%%%%%%%%%%%%%%%%%%%%%%%%%%%%%%%%%%%%%%%%%%%%%%%%%%%%%%%%
\section{Solar-type stars}
%%%%%%%%%%%%%%%%%%%%%%%%%%%%%%%%%%%%%%%%%%%%%%%%%%%%%%%%%%%%%%%%%%%%%%%%%%%%%
Global meridional flow in the convection zones of sun-like stars can be understood as a result of deviations from the thermal wind balance (Section 4). Figure 4 shows that considerable deviations can be found in relatively thin layers near the convection zone boundaries. Thickness of the (Ekman) layers decreases with the rotation rate of a star of a given mass. Computations with the differential rotation model, therefore, predict a decrease in the kinetic energy of the meridional flow with rotation rate (Kitchatinov and Olemskoy, 2012). However, the amplitude of the flow velocity in thin boundary layers increases. The layers in young stars rotating with a short period of about one day are so thin that the meridional flow probably cannot be significant for dynamos. The young stars dynamos are expected to differ from the solar case (Karak et al., 2014; Kitchatinov and Olemskoy, 2015). The solar-type flux-transport dynamo can have their onset in older stars with rotation periods of about ten days (Katsova et al., 2015).

The model computations also predict a moderate dependence on stellar mass. The meridional circulation is computed to be slower in smaller stars.

Baklanova and Plachida (2015) estimated the meridional flow for solar-type stars with known activity cycles. They assumed that the cycle period is controlled by the flow circulation time and find that a flow velocity of about 10\,m\,s$^{-1}$ is almost constant with the Rossby number.
%%%%%%%%%%%%%%%%%%%%%%%%%%%%%%%%%%%%%%%%%%%%%%%%%%%%%%%%%%%%%%%%%%%%%%%%%%%%%
\phantomsection
\section*{Acknowledgments}
\addcontentsline{toc}{section}{Acknowledgments} % Uncomment to add Acknowledgements to the table of contents
%%%%%%%%%%%%%%%%%%%%%%%%%%%%%%%%%%%%%%%%%%%%%%%%%%%%%%%%%%%%%%%%%%%%%%%%%%%%%
This work was supported by the Russian Foundation
for Basic Research (project 16--02--00090).
%%%%%%%%%%%%%%%%%%%%%%%%%%%%%%%%%%%%%%%%%%%%%%%%%%%%%%%%%%%%%%%%%%%%%%%%%%%%%
\phantomsection
\section*{References}
\addcontentsline{toc}{section}{References} %Uncomment to add References to content
%%%%%%%%%%%%%%%%%%%%%%%%%%%%%%%%%%%%%%%%%%%%%%%%%%%%%%%%%%%%%%%%%%%%%%%%%%%%%
\begin{description}
%%%%%%%%%%%%%%%%%%%%%%%%%%%%%%%%%%%%%%%%%%%%%%%%%%%%%%%%%%%%%%%%%%%%%%%%%%%%%
\item{} Antia, H.\,M., Basu, S., and Chitre, S.\,M., Solar rotation rate and its gradients during cycle 23, {\it Astrophys. J.}, 2008, vol.681, pp.680-692.
\item{} Baklanova, D., Plachinda, S., Meridional flow velocities on solar-like stars with known activity	cycles, {\it Adv. Space Res.}, 2015, vol.55, pp.817-821.
\item{} Choudhuri, A.R., Schüssler, M., and Dikpati, M., The solar dynamo with meridional circulation, {\it Astron. Astrophys.}, 1995, vol.303, pp.L29-L32.
\item{} Dicke, R.\,H., Internal rotation of the Sun, {\it Ann. Rev. Astron. Astrophys.}, 1970, vol.8, pp.297-328.
\item{} Durney, B.\,R., On the behavior of the angular velocity in the lower part of the solar convection zone, {\it Astrophys. J.}, 1989, vol.338, pp.509-527.
\item{} Durney, B.\,R., On a Babcock-Leighton dynamo model with a deep-seated generating layer for the toroidal magnetic field, {\it Solar Phys.}, 1995, vol.160, pp.213-235.
\item{} Duvall, T.\,L., Jefferies, S.\,M., Harvey, J.\,W., and Pomerantz, M.\,A., Time-distance helioseismology, {\it Nature}, 1993, vol.362, pp.430-432.
\item{} Einshtein, A., Die Ursache der M\"{a}anderbildung der Flu{\ss}l\"{a}ufe und das sogenannten Baerschen Gesetzes, {\it Die Naturwissenshaften}, 1926, vol.14, pp.223-224.
\item{} Featherstone, N.\,A. and Miesch, M.\,S., Meridional circulation in solar and stellar convection zones, {\it Astrophys. J.}, 2015, vol.804, id.:67.
\item{} Garaud, P. and Bodenheimer, P., Gyroscopic pumping of large-scale flows in stellar interiors and application to Lithium-deep stars, {\it Astrophys. J.}, 2010, vol.719, pp.313-334.
\item{} Giles, P.\,M., Duvall, T.\,L., Scherrer, P.\,H., and Bogart, R.\,S., A subsurface flow of material from the Sun’s equator to its poles, {\it Nature}, 1997, vol.390, pp.52-54.
\item{} Gilman, P.\,A. and Miesch, M.\,S., Limits to penetration of meridional circulation below the solar convection zone, {\it Astrophys. J.}, 2004, vol.611, pp.568-574.
\item{} Gough, D.\,O., Some glimpses from helioseismology at the dynamics of the deep solar interior, {\it Space Sci. Rev.}, 2015, vol.196, pp.15-47.
\item{} Guerrero, G., Smolarkiewicz, P.\,K., Kosovichev, A.\,G., and Mansour, N.\,N., Differential rotation in solar-like stars from global simulations, {\it Astropphys. J.}, 2013, vol.779, id.176.
\item{} Hathaway, D.\,H., and Rightmire, L., Variations in the Sun’s meridional flow over a solar cycle, {\it Science}, 2010, vol.327, pp.1350-1352.
\item{} Hazra, G., Karak, B.\,B., Choudhuri, A.\,R., Is a one-cell meridional circulation essential for the flux transport solar dynamo? {\it Astrophys. J.}, 2014, vol.782, id.: 93.
\item{} Jackewicz, J., Serebryanskiy, A., and Kholikov, S., Meridional flow in the solar convection zone. II. Helioseismic inversions of GONG data, {\it Astrophys. J.}, 2015, vol.805, id.: 133.
\item{} Käpylä, P.\,J., Mantere, M.\,J., Guerrero, G., Brandenburg, A., and Chatterjee, P., Reynolds stress and heat flux in spherical shell convection, {\it Astron. Astrophys.}, 2011, vol.531, id.: A162.
\item{} Karak, B.\,B. and Choudhuri, A.\,R., A possible explanation of the Maunder minimum from a flux transport dynamo model, {\it Research in Astron. Astrophys.}, 2009, vol.9, pp.953-958.
\item{} Karak, B.\,B., Kitchatinov, L.\,L., and Choudhuri, A.\,R., A dynamo model of magnetic activity in solar-like stars with different rotational velocities, {\it Astrophys. J.}, 2014, vol.791, id.: 59.
\item{} Karak, B.\,B., Käpylä, P.\,J., Käpylä, M.\,J., Brandenburg, A., Olspert, N., and Pelt, J., Magnetically controlled stellar differential rotation near the transition from solar to anti-solar profiles, {\it Astron. Astrophys.}, 2015, vol.576, id.: A26.
\item{} Katsova, M.\,M., Bondar, N.\,I., and Livshits, M.\,A., Solar-type activity: Epochs of cycle formation, {\it Astron. Rep.}, 2015, vol. 59, pp.726-735.
\item{} Kippenhahn, R., Differential rotation in stars with convective envelopes, {\it Astrophys. J.}, 1963, vol.137, pp.664-678.
\item{} Kitchatinov, L.\,L., Differential rotation of a star induced by meridional circulation, {\it Astron. Rep.}, 2006, vol.50, pp.512-516.
\item{} Kitchatinov, L.\,L., Theory of differential rotation and meridional circulation, in Proc. IAU Symp. 294 {\it \lq Solar and Astrophysical Dynamos and Magnetic Activity'}, Kosovcichev, A.G., de Gouveia Dal Pino, E., Yan, Y., Eds., Cambridge Univ. Press, 2013, pp.399-410.
\item{} Kitchatinov, L.\,L. and Rüdiger, G., Differential rotation in solar-type stars: revisiting the Taylor-number puzzle, {\it Astron. Astrophys.}, 1995, vol.299, pp.446-452.
\item{} Kitchatinov, L.\,L. and Rüdiger, G., Differential rotation in the solar convection zone and beneath, {\it Astron. Nachr.}, 2005, vol.326, pp.379-385.
\item{} Kitchatinov, L.\,L. and Olemskoy, S.\,V., Differential rotation of main-sequence dwarfs and its Dynamo efficiency, {\it Mon. Not. Roy. Astron. Soc.}, 2011, vol.411, pp.1059-1066.
\item{} Kitchatinov, L.\,L. and Olemskoy, S.\,V., Differential rotation of main-sequence dwarfs: predicting the dependence on surface temperature and rotation rate, {\it Mon. Not. Roy. Astron. Soc.}, 2012, vol.423, pp.3344-3351.
\item{} Kitchatinov, L.\,L. and Olemskoy, S.\,V., Dynamo saturation in rapidly rotating solar-type stars, {\it Research in Astron. Astrophys.}, 2015, vol.15, pp.1801-1812.
\item{} Komm, R.\,W., Howard, R.\,F., Harvey, J.\,W., Meridional flow of small photospheric magnetic features, {\it Sol. Phys.}, 1993, vol.147, pp.207-223.
\item{} Kuhn, J.\,R., Bush, R., Emilio, M., Scholl, I.\,F., The precise solar shape and its variability, {\it Science}, 2012, vol.337, pp.1638-1640.
\item{} Matt, S.\,P., Do Chao, O., Brown, B.\,P., and Brun, A.\,S., Convection and differential rotation properties of G and K stars computed with the ASH code, {\it Astron. Nachr.}, 2011, vol. 332, pp.897-907.
\item{} Miesch, M.\,S., Brun, A.\,S.,  and Toomre, J., Solar differential rotation influenced by latitudinal entropy variations in the tachocline, {\it Astrophys. J.}, 2006, vol.641, pp.618-625.
\item{} Passos, D., Charbonneau, P., Miesch, M., Meridional circulation dynamics from 3D magnetohydrodynamic global simulations of solar convection, {\it Astrophys. J.}, 2015, vol.800, id.: L18.
\item{} Rajaguru, S.\,P. and Antia, H.\,M., Meridional circulation in the solar convection zone: Time-distance helioseismic inferences from four years of HMI/SDO observations, {\it Astrophys. J.}, 2015, vol.813, id.: 114.
\item{} Rast, M.\,P., Ortiz, A., and Meisner, R.\,W., Latitudinal variation of the solar photospheric intensity, {\it Astrophys. J.}, 2008, vol.673, pp.1209-1217.
\item{} Rempel, M., Influence of random fluctuations in the $\Lambda$-effect on meridional flow and differential rotation, {\it Astrophys. J.}, 2005, vol.631, pp.1286-1292.
\item{} R\"{u}diger, G., {\it Differential rotation and Stellar Convection}, New York: Gordon \& Breach, 1989.
\newpage
\item{} R\"{u}diger, G., Egorov, P., Kitchatinov, L.\,L., and K\"{u}ker, M., The eddy heat-flux in rotating turbulent	convection, {\it Astron. Astrophys.}, 2005, vol.431, 345-352.
\item{} R\"{u}diger, G., Kitchatinov, L.\,L., and Hollerbach, R., {\it Magnetic Processes in Astrophysics}, Weinheim:	Wiley-VCH, 2013.
\item{} Shad, A., Timmer, J., and Roth, M., Global helioseismic evidence for a deeply penetrating solar meridional flow consisting of multiple flow cells, {\it Astrophys. J.}, 2013, vol.778, id.: L38.
\item{} Sheeley, N.\,R., Jr., Surface evolution of the Sun’s magnetic field: A hystorical review of the flux-transport mechanism, {\it Living Rev. Solar Phys.}, 2005, vol. 2, 27 pp.
\item{} Tassoul, J.-L., {\it Stellar Rotation}, Cambridge University Press, 2000.
\item{} Wang, Y.-M., Sheeley, N.\,R., Jr., and Nash, A.\,G., Anew solar cycle model including meridional circulation, {\it Astrophys. J.}, 1991, vol.383, pp.431-442.
\item{} Warnecke, J., Käpylä, P.\,J., Mantere, M.\,J.,  and Brandenburg, A., Spoke-like differential rotation in a convective dynamo with a coronal envelope, {\it Astrophys. J.}, 2013, vol.778, id.:141.
\item{} Weiss, N.\,O., Convection and the differential rotation of the Sun, {\it Observatory}, 1965, vol.85, pp.37-39.
\item{} Zhao, J. and Kosovichev, A.\,G., Detection of equatorward meridional flow and evidence of double-cell meridional circulation inside the Sun, {\it Astrophys. J.}, 2013, vol.774, id.: L29.
\end{description}
\end{document}